\PassOptionsToPackage{table}{xcolor}
\PassOptionsToPackage{footnotesize}{caption}

\documentclass[10pt,conference]{IEEEtran}
\usepackage[letterpaper, left=1in, right=1in, bottom=1in, top=0.75in]{geometry}

\usepackage{amsmath,amssymb,amsfonts}
\usepackage{setspace}
\usepackage{libertine}
\usepackage{libertinust1math}
\usepackage[T1]{fontenc}
\usepackage{inconsolata}
\IfFileExists{microtype.sty}{
  \usepackage[]{microtype}
  \UseMicrotypeSet[protrusion]{basicmath} % disable protrusion for tt fonts
}{}
\usepackage{xcolor}
\IfFileExists{xurl.sty}{\usepackage{xurl}}{}

\usepackage{geometry}
\usepackage{scrlayer-scrpage}
\usepackage{longtable,booktabs,array,tabularx}
\usepackage{paralist}
\usepackage{pdfpages}
\usepackage{subcaption}
\usepackage{xspace}
\usepackage{multirow}
\usepackage{calc}
\usepackage{markdown}
\usepackage{graphicx}
\usepackage[normalem]{ulem}
\usepackage{enumitem}
\setlist[enumerate]{noitemsep}
\setlist[itemize]{noitemsep}

\usepackage{listings}
\usepackage[cache=true,frozencache,newfloat=true]{minted}
\setminted{fontsize=\scriptsize}

\usepackage{csquotes}

\usepackage{makecell} % For table 3 - newlines in cells

\newcommand{\shorttitle}{Homo in Machina\xspace}

\newcommand{\ccode}[1]{\mintinline{c}|#1|}
\newcommand{\aflpp}[0]{\texttt{AFL++}\xspace}

\newcommand{\file}[0]{\texttt{file}}
\newcommand{\cone}[0]{\raisebox{.5pt}{\textcircled{\raisebox{-.9pt} {1}}}}
\newcommand{\ctwo}[0]{\raisebox{.5pt}{\textcircled{\raisebox{-.9pt} {2}}}}
\newcommand{\cthree}[0]{\raisebox{.5pt}{\textcircled{\raisebox{-.9pt} {3}}}}
\newsavebox{\mintedbox}

\lstdefinestyle{mySmallPython}{%
    language=Python,
    tabsize=4
}

\usepackage{hyperref}
\hypersetup{%
    pdftitle=\shorttitle,
    pdfauthor=Anonymized for Submission,
    colorlinks=true,
    linkcolor={red},
    filecolor={red},
    citecolor={blue},
    urlcolor={blue}}
\usepackage[backend=biber,style=ieee,sorting=ynt,maxbibnames=2,minbibnames=1,doi=false,url=false,isbn=false,eprint=false]{biblatex}
\def\BibTeX{{\rm B\kern-.05em{\sc i\kern-.025em b}\kern-.08em
    T\kern-.1667em\lower.7ex\hbox{E}\kern-.125emX}}

\title{\emph{Homo in Machina\xspace}: Improving Fuzz Testing Coverage via Compartment Analysis}

\author{%
    \IEEEauthorblockN{%
        Joshua Bundt\IEEEauthorrefmark{1}\IEEEauthorrefmark{4},
        Andrew Fasano\IEEEauthorrefmark{1}\IEEEauthorrefmark{2},
        Brendan Dolan-Gavitt\IEEEauthorrefmark{3},
        William Robertson\IEEEauthorrefmark{1},
        Tim Leek\IEEEauthorrefmark{2}%
        }%
\IEEEauthorblockA{%
    \IEEEauthorrefmark{1}\textit{Northeastern University},
    \IEEEauthorrefmark{3}\textit{New York University}}%
\IEEEauthorblockA{%
    \IEEEauthorrefmark{4}\textit{Army Cyber Institute},
    \IEEEauthorrefmark{2}\textit{MIT Lincoln Laboratory}}%
}

\date{}

\begin{document}

\maketitle

\begin{abstract}
Fuzz testing is often automated, but also frequently augmented by experts who insert themselves into the workflow in a greedy search for bugs.
In this paper, we propose \emph{Homo in Machina}, or HM-fuzzing, in which analyses guide the manual efforts, maximizing benefit.
As one example of this paradigm, we introduce \emph{compartment analysis}.
Compartment analysis uses a whole-program dominator analysis to estimate the utility of reaching new code, and combines this with a dynamic analysis indicating drastically under-covered edges guarding that code.
This results in a prioritized list of \emph{compartments}, i.e., large, uncovered parts of the program semantically partitioned and largely unreachable given the current corpus of inputs under consideration.
% A dynamic data flow analysis is further used to sort the conditionals guarding compartments into bins: those dependent on user input, those controlled by the fuzzing harness, and those seemingly untainted ones requiring manual analysis.
A human can use this categorization and ranking of compartments directly to focus manual effort, finding or fashioning inputs that make the compartments available for future fuzzing.
We evaluate the effect of compartment analysis on seven projects within the OSS-Fuzz corpus where we see coverage improvements over AFL++ as high as 94\%, with a median of 13\%.
We further observe that the determination of compartments is highly stable and thus can be done early in a fuzzing campaign, maximizing the potential for impact.

\end{abstract}

\section{Introduction}%
\label{sec:introduction}

Fuzz testing, or fuzzing, is the most effective automated vulnerability discovery technique available today.  It is widely used in industry~\cite{godefroid_2012_sagewhiteboxfuzzing,google_2020_clusterfuzz,google_2020_ossfuzz,google_fuzzbench}, and has been the subject of intense research in recent years~\cite{manes_2019_artscienceengineering}.  However, as a dynamic testing technique, fuzzing is fundamentally limited---if code is not executed, then fuzzing cannot discover bugs within that code.
Coverage, usually measured in terms of basic blocks or edges that have been exeucted at least once, is a necessary condition for surfacing unknown bugs.  Note that it is not a \emph{sufficient} condition; just because a block containing a bug has been executed does not mean that bug will be triggered without satisfying additional constraints on program state in the general case.  However, achieving a high coverage rate remains a vitally important hurdle to overcome when using fuzzing to discover new vulnerabilities.

Unfortunately, fuzzing is known to be unable to cover most of a program under test even when using simple block or edge coverage metrics.  Empirical studies have repeatedly shown that coverage tends to plateau relatively quickly, usually within 24~hours~\cite{klees_2018_evaluatingfuzztesting,bundt_2021_evaluatingsyntheticbugs}.  Recent work examined this phenomenon from a theoretical perspective, postulating an empirical power law governing the difficulty of achieving new coverage---new coverage becomes exponentially more difficult to obtain over time~\cite{bohme_2020_fuzzingexponentialcost}.
Numerous approaches have been proposed, in part or in full, to surmount this ``coverage barrier'':
\begin{inparaenum}[\itshape(i)\upshape]
    \item seed selection and trimming to reduce the number of wasteful trials on inputs that are unlikely to produce new coverage~\cite{woo_2013_schedulingblackboxmutational,rebert_2014_optimizingseedselection};
    \item more sophisticated fuzz configuration scheduling algorithms~\cite{bohme_2016_coveragebasedgreyboxfuzzing,bohme_2017_directedgreyboxfuzzing};
    \item improving mutation operator scheduling~\cite{lyu_2019_moptoptimizedmutation};
    \item input region prioritization~\cite{ganesh_2009_taintbaseddirectedwhitebox,wang_2010_taintscopechecksumawaredirected,rawat_2017_vuzzerapplicationawareevolutionary,chen_2018_angoraefficientfuzzing,aschermann_2019_redqueenfuzzinginputtostate};
    \item hybrid fuzzing approaches that use concolic execution to assist a conventional grey-box mutational fuzzer~\cite{stephens_2016_drilleraugmentingfuzzing,yun_2018_qsympracticalconcolic,zhao_2019_sendhardestproblems}; and
    \item learning from human input~\cite{shoshitaishvili_2017_risehacrsaugmenting}, among others.
\end{inparaenum}
Despite this enormous research investment, improving fuzzers to cover new code over time and across individual fuzzing campaigns remains an open and central problem in the field.

In this work, we propose \emph{Homo in Machina} or HM-fuzzing, which presents conclusions derived from incremental fuzzing results to a human who can in turn help guide the fuzzer to achieve greater coverage than would otherwise be possible on its own.  Currently, humans primarily intervene in fuzzing campaigns by analyzing coverage reports and either providing new seeds or modifying test harnesses.  HM-fuzzing aims to improve this critical, yet understudied, manual aspect of real-world fuzzing, building on the key insight that while humans are substantially more capable than automated techniques at covering new code, they also have a relatively limited budget to apply this expertise.

As one example of HM-fuzzing, we introduce \emph{compartment analysis}.  Compartment analysis is designed to be run periodically during a fuzzing campaign on the incremental coverage achieved to that point.  The goal of the analysis is to identify \emph{compartments}, or code that is dominated by a block adjacent to the coverage frontier~\cite{leek_2007_coveragemaximizationusing} that, if covered, would likely maximize overall coverage given a fixed resource budget.  The interprocedural control-flow graph (ICFG) of a program under test is used to identify nodes that dominate the greatest number of basic blocks when considering both intra- and interprocedural edges.  These compartments are further filtered according to how \emph{saturated} they are given the current fuzzing corpus, retaining only severely under-covered or entirely uncovered compartments.  A dynamic data-flow analysis is then used in conjunction with the ICFG to provide additional advice to the analyst about how the conditionals that gate these compartments depend upon inputs or the fuzzing test harness.  The analysis orders these compartments with a weighting function, and returns the heavy hitters to a security analyst.  The analyst can then use signals extracted from the compartments to adjust the test harness or augment the input queue to ``unlock'' those compartments.  This technique will not only improve coverage for the remainder of a fuzzing campaign, but also for \emph{future} campaigns as well.
Compartment analysis systematically enables a security analyst to judiciously intervene to assist fuzzers in covering code that they likely could not cover on their own.

Our evaluation applies a compartment analysis to a test set of programs and corresponding fuzzing harnesses from the Google FuzzBench corpus~\cite{google_fuzzbench}.
The overall results demonstrate that compartment analysis is effective at obtaining substantial improvements in coverage, with gains of at least 10\% in six of seven cases and an increase of 94\% for one.
Human effort was estimated at several hours per program, which is modest given the gains.

In the interests of open science and reproducibility, we will open-source our prototype implementation and data after publication.\footnote{An anonymous code and data repository is provided at~\url{https://osf.io/7jwv5/?view_only=c37c8c6789934407a3b85c69f2a8743e}.}
In summary, our paper makes the following contributions:
\begin{itemize}
    \item We propose \emph{compartment analysis} as a novel human intervention in fuzz testing to significantly improve coverage for current \emph{and} future campaigns.
    \item We implement a prototype of this analysis that suggests rank-ordered interventions in terms of testing harness modifications and input seed set augmentation.
    \item We evaluate our prototype on elements of the Google FuzzBench corpus and demonstrate significant improvements in coverage over baseline campaigns.
\end{itemize}

\begin{figure}[tb]
    \centering
    \includegraphics[width=0.7\linewidth]{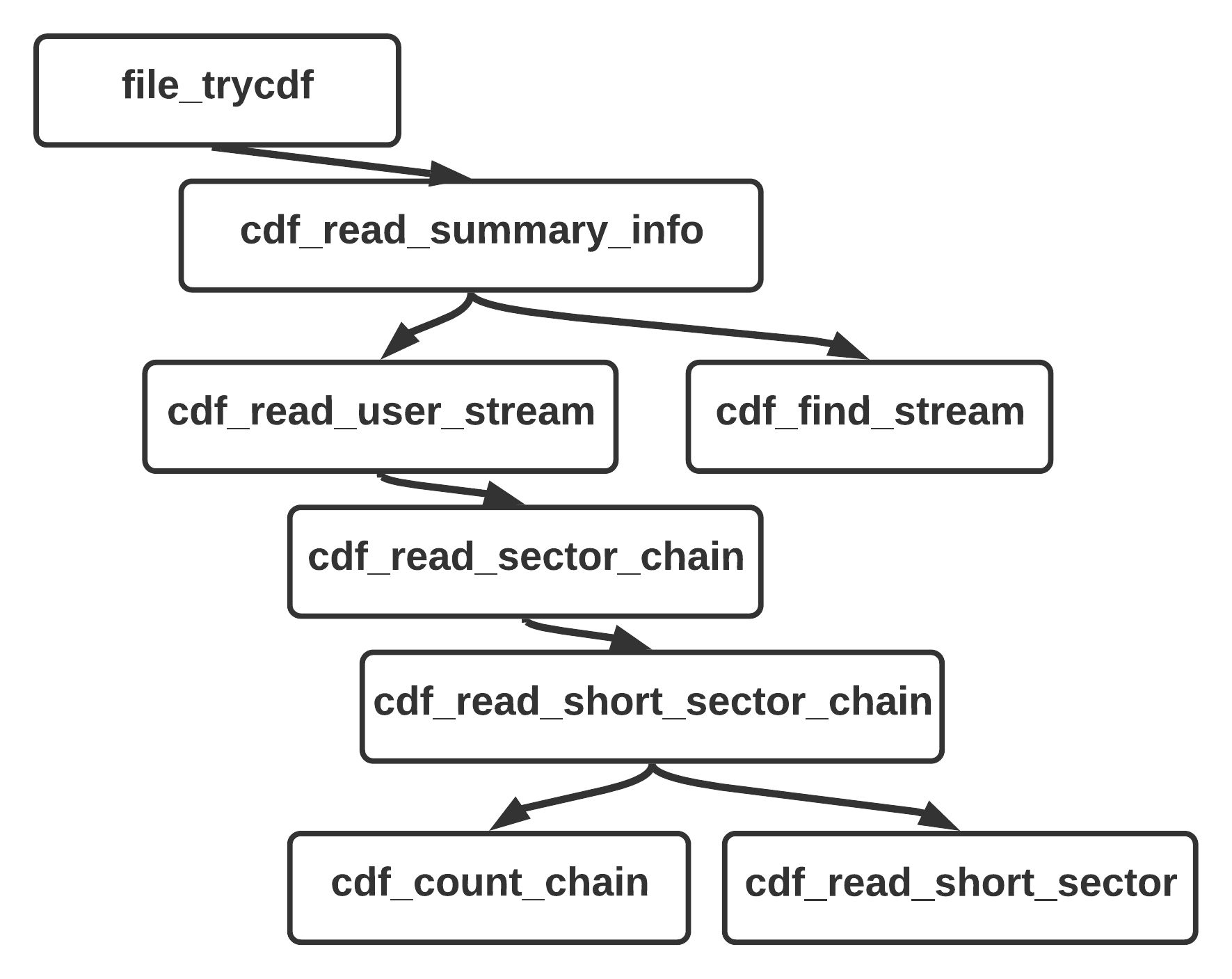}
    \caption{Call graph for compartment \texttt{file\_trycdf: \%114}.  A coverage guided fuzzer must synthesize an input that will satisify the conditions of the full call-chain  in order to explore the CDF parsing code in \file.}%
    \label{figure:file-trycdf-callgraph}
\end{figure}

\section{Motivation and Problem Statement}%
\label{sec:motivation}

\begin{figure*}[t]
    \centering
    \includegraphics[width=1.00\textwidth]{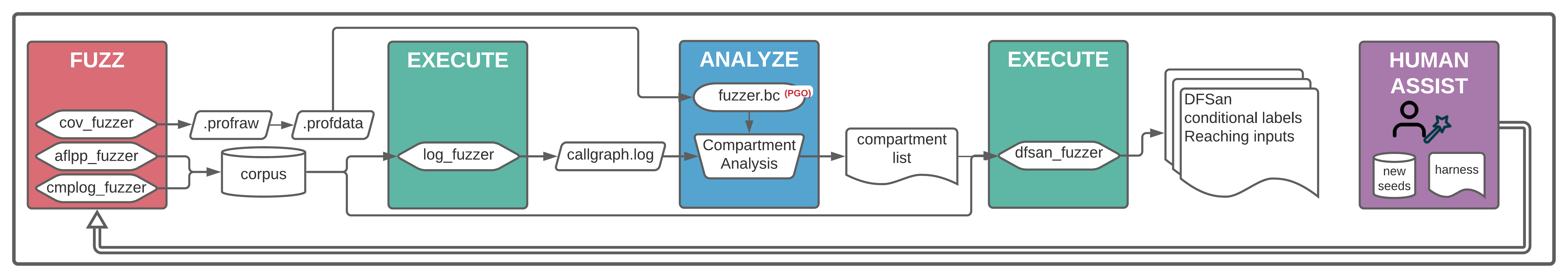}
    \caption{Overview of system architecture and workflow.}%
    \label{fig:architecture}
\end{figure*}

The ultimate goal of this work is to assist fuzzers in covering code that they would otherwise be unable to reach and saturate during a fuzzing campaign.  However, it is important to recognize that there are multiple reasons why code might be difficult to cover.
The most obvious reasons that coverage-guided fuzzers get stuck is due to complex program logic that does not lend itself to incremental progress or reduction to satisfiable formulas.
Alternatively, a fuzzer maybe prevented from covering code due to the choices made in how the fuzzer will load and execute the program under test.

\noindent{\bf Problem:} Consider Figure~\ref{figure:file-trycdf-callgraph}, which shows a call subgraph for the ``file'' utility rooted at \ccode{file_trycdf}.  This particular code is responsible for identifying and summarizing a Microsoft Composite Document File (CDF).\footnote{\url{http://sc.openoffice.org/compdocfileformat.pdf}}  Covering code that specifically handles CDFs is exceedingly difficult for state-of-the-art fuzzers without an example of such a file provided as a seed.  A CDF is a highly structured and complicated file format that essentially encodes a FAT-style file system within a file.  To explore this code without a valid example, a fuzzer would need to mutationally synthesize a valid CDF file from some other seed.  While fuzzers have famously been observed to automatically synthesize instances of some file formats~\cite{zalewski_2014_pullingjpegsout}, it is highly unlikely that a fuzzer would be able to pull off this feat for highly structured inputs like CDFs.

From the fuzzer's perspective, mutational synthesis of a CDF corresponds to satisfying a complex symbolic formula representing the path constraints leading to \emph{and into} file's CDF code shown in Figure~\ref{figure:file-trycdf-callgraph}.  Satisfying complex path constraints, even with the help of specialized tools engineered exactly for this purpose such as an SMT solver, is well known to be difficult for fuzzers, and is a large contributor to the fuzz testing coverage problem.

\begin{listing}[tb]
    \begin{lrbox}{\mintedbox}
    \RecustomVerbatimEnvironment{Verbatim}{BVerbatim}{}
    \begin{minted}{diff}
  extern "C" int LLVMFuzzerTestOneInput(const
                    uint8_t *data, size_t size) {%
+    int options = XML_PARSE_DTDVALID;
     xmlSetGenericErrorFunc(NULL, &ignore);
     if (auto doc = xmlReadMemory(reinterpret_cast<const char *>(data),
-                                 size, NULL, 0))
+                                 size, NULL, options))
+        xmlFree(doc);
+    options = XML_PARSE_SAX1;
+    if (auto doc = xmlReadMemory(reinterpret_cast<const char *>(data),
+                                 size, NULL, options))
         xmlFree(doc);
     return 0;
 }
    \end{minted}
    \end{lrbox}
    \resizebox{\linewidth}{!}{\usebox{\mintedbox}}
    \caption{Motivational example diff showing how a libFuzzer-based test harness can be modified to reach previously uncovered compartments.}
    \label{lst:example_libxml2}
\end{listing}

\noindent{\bf Problem:} On the other hand, consider Listing~\ref{lst:example_libxml2}, which shows a diff of a snippet of code from a libFuzzer harness for the libxml2 XML parsing library.  In this case, we can see that the original version of the harness passes in an empty \ccode{options} bitset.  While this might seem innocuous, in practice the effect of this choice is to deterministically exclude from all fuzzing trials all code in libxml2 that handles options indicated by that parameter to \ccode{xmlReadMemory}.  In other words, because \ccode{options} is set to a fixed value by the harness, libFuzzer will \emph{never} cover a sizeable subset of libxml2.  Note that this problem is not limited to libFuzzer, but rather generalizes to whole program testing where the command line or environmental variables are either initialized to a default value or not mutated.

The vast majority of prior work has focused on addressing the first of these two issues, in which a new type or class of input is required to address the coverage problem~\cite{woo_2013_schedulingblackboxmutational,rebert_2014_optimizingseedselection,bohme_2016_coveragebasedgreyboxfuzzing,bohme_2017_directedgreyboxfuzzing,lyu_2019_moptoptimizedmutation,ganesh_2009_taintbaseddirectedwhitebox,wang_2010_taintscopechecksumawaredirected,rawat_2017_vuzzerapplicationawareevolutionary,chen_2018_angoraefficientfuzzing,aschermann_2019_redqueenfuzzinginputtostate,stephens_2016_drilleraugmentingfuzzing,yun_2018_qsympracticalconcolic,zhao_2019_sendhardestproblems,shoshitaishvili_2017_risehacrsaugmenting}.
We are not aware of work that has systematically addressed the second issue, in which coverage gains are effectively blocked, by a missing \emph{source} input, be it a command-line option, an environment variable, or a host of other unanticipated input sources.

Addressing these issues, when automated means fail, is the job of the human analyst.
To enable a fuzzer to reach new CDF parsing code in file, an analyst might
peruse the relevant portions of uncovered source code, determine that this kind of input is missing from the corpus, and obtain a few new seeds with some Google searches.
To fuzz the SAX1 interface of libxml2, an analyst might dive into the source code of the target program to find that option-specific code in libxml2 is not being covered. With this knowledge in hand, a brief survey of the harness code will suggest a fix.
In both cases, these are tasks for experts who need to understand what the program does, how it is implemented, how its inputs are structured, and how to find new seeds.
These are good tasks for the human analyst who can integrate the diverse skills into a holistic problem-solving workflow easily.

% Unfortunately, coverage remains a central obstacle for automated fuzz testing, and it is ultimately left to the security analyst to recognize these problems and address them manually.  For example, one might notice that the CDF code in file had not been explored during a campaign, and optimistically add a CDF seed to the input queue.  Or, one might recognize that option-specific code in libxml2 had not been covered, trace this back to the fuzzing harness, and modify it to explore different values for \ccode{options} as shown in Listing~\ref{lst:example_libxml2}.

While security analysts orchestrating fuzzing campaigns do perform this sort of manual analysis, there is little guidance available to them.
Yet this sort of targeted human intervention can be extremely effective in assisting automated fuzzers.
Thus, we pose the following research question.

\vspace{0.1in}
\noindent
\emph{\underline{Research Question}: Can fuzzers systematically provide this missing guidance, and suggest high value interventions to a security analyst in the form of input queue augmentation or harness modifications, to improve automated fuzz testing coverage?}

\setlength{\tabcolsep}{2pt} % Very reduced padding for this table
\begin{table*}
\centering
\scriptsize
\begin{tabularx}{\linewidth}{rXrrrrcllll}
\toprule
\textbf{Rank} & \textbf{Function} & \textbf{Weight}  & \textbf{Block Weight} & \textbf{Calls Weight} & \textbf{Profile Cnt} & \textbf{Label}\cone{} & \textbf{Conditional} & \textbf{Compartment} & \textbf{Input}\ctwo{} & \textbf{Solution}\cthree{} \\
\midrule
  1 & pfr\_face\_init  &       2241 &    482 &    1759 & 427023610 &    & pfrobjs.c:88   & pfrobjs.c:97      & Zurich.pfr        & Zurich.pfr       \\
  2 & pcf\_load\_font  &       1242 &    603 &     639 &    442315 &    & pcfread.c:1376 & pcfread.c:1380    & courB10.pcf.Z     & courB10.pcf.Z    \\
  3 & cid\_face\_open  &        968 &    275 &     693 &   4683065 &    & cidload.c:716  & cidload.c:719     & 96h\_004325       &                  \\
  4 & woff\_open\_font &        943 &    850 &      93 &      8487 & I  & sfobjs.c:452   & sfobjs.c:453      & Lack.woff         & FiraCode-VF.woff \\
\bottomrule
\end{tabularx}
    \caption{Example compartment list extracted from freetype2. \textbf{Weight}: total weight. \textbf{Profile Cnt}: execution count at conditional. \textbf{Label}: I=input label. \textbf{Conditional}: source location of conditional blocking compartment. \textbf{Compartment}: source location of entry block to compartment. \textbf{Input}: input file which reaches conditional. \textbf{Solution}: input file that covers entry block of compartment. Adding a Portable Font Resource (PFR) font such as Zurich.pfr will `unlock' the first compartment which has the greatest potential for new coverage.}%
\label{table:example_file_compartment_list}
\end{table*}

\section{Compartment Analysis}%
\label{sec:design}

The compartment analysis workflow is depicted in Figure~\ref{fig:architecture}.
In the first, or \verb+FUZZ+ step, the target program is subject to fuzz testing on some seed corpus, during which edge counts are accumulated across all mutated inputs and output as LLVM profiling data files (\verb+*.profraw+) at the end of the campaign, along with a corpus of coverage-improving or crash-inducing inputs, i.e.\ the fuzzing ``queue.''
The queue is then re-executed to obtain a dynamic call graph in the first \verb+EXECUTE+ step, in order to collect indirect call edges traversed in \verb+callgraph.log+.
This dynamic indirect call data, along with edge counts, are used in the \verb+ANALYZE+ step to identify compartments and compute weights as described in Section~\ref{sub:optimal_edge_ranking}.
The output of this step is a \verb+compartment list+ which is run through another \verb+EXECUTE+ step, in which a \verb+dfsan+ instrumented version of the program is run to diagnose compartment-guarding conditions according to what source of input controls them (see Section~\ref{sub:solution_hints}).
This results in a compartment ranking report similar to the example in Table~\ref{table:example_file_compartment_list}, which a human will use to guide her effort, taking the form of adding new seeds to the fuzzing queue or tinkering with the harness to expose new inputs to mutation.

\subsection{Coverage Collection}%
\label{sub:coverage_collection}

Collecting coverage data is an integral part of fuzzing, especially for the most popular form of fuzzing: coverage-guided fuzzing.
The methods for collecting and storing coverage information during fuzzing have been thoroughly studied%
~\cite{wang_2019_coverage_metrics,gan_2018_collafl,wang_2020_coverage_accounting,yan_2020_PathAFL}%
, with efforts focusing on selecting the right level of sensitivity and efficiency for identifying new coverage.
Although there are some distinctions in techniques, generally coverage-guided fuzzers keep track of what blocks or edges have been covered, and generate a mapping between the saved inputs in the corpus and their contribution to overall coverage~\cite{manes_2019_artscienceengineering}.

In a traditional fuzzing workflow, a researcher might review the progress a fuzzer has made by generating a coverage report.
Both GCC and Clang support instrumenting a binary and generating an HTML report of source-based code coverage in terms of \emph{code regions}, a metric that is more precise than line numbers.\footnote{\url{https://clang.llvm.org/docs/SourceBasedCodeCoverage.html}}
This report is generated by compiling an instrumented binary and executing the binary on the fuzzing corpus.
While this report will highlight functions with limited or zero coverage, it does not indicate the balance of resources the fuzzer has spent on various code regions.
In order to analyze code regions that might be under-explored, more detailed profiling information must be collected during fuzzing.
To support collecting profiling information during a fuzzing campaign, \aflpp~was modified by adding an additional fork-server (\texttt{cov\_fuzzer} in Figure~\ref{fig:architecture}).
This fork-server simply executes every modified input generated by the fuzzer on the instrumented binary.
The fuzzing slow-down is minimal, as profiling is collected in parallel with the default fork-server, but the compute cost is roughly doubled because the program is run twice.

\subsection{Compartment Ranking}%
\label{sub:optimal_edge_ranking}

\begin{listing}[t]
    \begin{minted}[]{python}
def getCallsWeight(function, visited):
    if function in visited:
        return 0
    visited.add(function)
    if function.profileEntryCount > MAX_EXEC_COUNT or \
       length(function.getCallers()) > 1:
        return 0
    weight = function.size
    for call in function.getCalls():
        callee = call.getCallee()
        weight += getCallsWeight(callee, visited)
    return weight

def getBlockWeight(domTree, basicBlock):
    if basicBlock.profileCount > MAX_EXEC_COUNT:
        return 0
    Descendents = domTree.getDescendents(basicBlock)
    CalledFunctions = basicBlock.getCalledFunctions()
    weight = basicBlock.size
    for domBlock in Descendents:
        weight += domBlock.size
        CalledFunctions.extend(domBlock.getCalledFunctions())
    visited = set()
    for function in CalledFunctions:
        weight += getCallsWeight(function, visited)
    return weight
    \end{minted}
    \caption{Compartment weighting algorithm pseudocode. \texttt{getBlockWeight} returns the total weight for a compartment which uses the recursive function \texttt{getCallsWeight} to determine the sum of uncovered code in uniquely reachable functions.}
    \label{lst:compartment_pseudocode}
\end{listing}

Compartments are ranked, or prioritized, based on the upper bound of reward in unlocking the compartment.
During a fuzzing campaign, the reward is additional new coverage.
The rank is determined by static analysis of the interprocedural control-flow graph (ICFG) combined with profiling coverage information attained during initial fuzzing.
The functions that determine the weight of a compartment are outlined in Listing~\ref{lst:compartment_pseudocode}.
Function \texttt{getBlockWeight} takes as inputs a dominator tree \(\mathbb{DT}\) derived from the ICFG and a basic block \(\mathbb{B}\) to analyze and returns a total weight.
In order to determine the instructions reached beyond function calls, \texttt{getCallsWeight} traverses the call graph in depth-first order, summing the weight for all functions that are uniquely reachable from call sites in \(\mathbb{DT}\).
The total weight of a compartment is the sum of LLVM intermediate representation (IR) instructions that are uniquely reachable from the entry block of the compartment.

% The function \texttt{getBlockWeight} takes as inputs a dominator tree \(\mathbb{DT}\) for a given function \(\mathbb{F}\) and a basic block \(\mathbb{B}\) to analyze and returns a total weight.
% The basic block used as input is the entry block to the compartment.
% The dominator tree method \texttt{getDescendents} returns all blocks dominated by \(\mathbb{B}\) according to the traditional dominance relationship where block \(d\) dominates block \(b\) in \(F\) if every path from the entry point to \(b\) must first traverse through \(d\).
% In order to determine code that could be reached through function calls, the basic block method \texttt{getCalledFunctions} returns a list of the functions called by a basic block.

% The function \texttt{getCallsWeight} takes as inputs an initial function \(\mathbb{F}\) and an empty working set.
% The function method \texttt{getCalls} returns a list of all function calls contained in \(\mathbb{F}\).
% \texttt{getCallsWeight} traverses the call graph in depth-first order, summing the weight for all functions that are uniquely reachable starting from \(\mathbb{F}\).
% The size of a block or function is merely the number of instructions that belong to each block or function, respectively.

A user defined parameter \texttt{MAX\_EXEC\_COUNT} allows the analyst to review unexecuted code and/or under-saturated code if desired.
For the purposes of evaluation, we chose to merely review compartments that were not covered during fuzzing.
Therefore, a low value for \texttt{MAX\_EXEC\_COUNT} (i.e., 50) is appropriate.
However, in a continuous fuzzing environment an analyst might be interested in compartments that are not adequately fuzzed, or under-saturated by fuzzing resources.
In this case, an analyst might set \texttt{MAX\_EXEC\_COUNT} to a higher value that captures the relative amount of under-saturation they wish to consider.

\begin{table}
\footnotesize
\centering
\begin{tabularx}{\linewidth}{Xrrr}
\toprule
 \textbf{Benchmark} & \textbf{Total} & \textbf{Indirect} & \textbf{Discovered} \\
\midrule
 file          &    3604 &          3 &            0 \\
 freetype2     &   16998 &        654 &          336 \\
 lcms          &    7649 &        594 &          123 \\
 libjpeg-turbo &   14423 &        603 &          254 \\
 libpcap       &    3895 &         14 &           11 \\
 libxml2       &    9328 &       1027 &          310 \\
 proj4         &   10413 &        495 &           84 \\
\bottomrule
\end{tabularx}
    \caption{Breakdown of total call sites, indirect call sites, and unique indirect call targets discovered after 96~hours of fuzzing. The amount of code `hidden' behind indirect calls varies greatly between benchmarks.}%
    \label{table:indirect-call-numbers}
\end{table}

\subsection{Dynamic ICFG Considerations}%
\label{sub:dynamic_interprocedural_cfg_construction}

While much of the ICFG can be determined at compile time, indirect call sites prevent precise whole program analysis.
There are tools available~\cite{lehr_2020_metacg,schubert_2019_phasar} to perform call graph analysis, but they tend to over-approximate the call graph edges at indirect call sites.
This is due to the overlap in function prototypes, especially in C programs, which are the primary means of determining indirect call edges.
For instance in \texttt{PROJ4} the function \texttt{pj\_init\_ctx} contains several indirect call-sites, one which takes a single pointer argument and returns a pointer.
In the analysis of our chosen benchmarks, indirect calls sites can account for as much as 11\% of the total number of nodes within the ICFG and a greater percentage of edges.
Additionally, many indirect calls can still remain undiscovered even after four days of fuzzing as shown in Table~\ref{table:indirect-call-numbers}.
We chose not to rely on static analysis to resolve indirect calls, as false alarms are common enough to either compromise human trust in tool output or reduce workflow efficiency.
In programs where much of the uncovered code lies behind indirect call targets, note that we can simply consider each function without any incoming direct call edges as a possible compartment, and compute its weight as $W_{\theta}(b)$, where $b$ is the entry basic block of the function.

\subsection{Compartment Solution Hints}%
\label{sub:solution_hints}

The LLVM DataFlowSanitizer\footnote{https://clang.llvm.org/docs/DataFlowSanitizer.html} (DFSan) is a powerful dynamic data flow analysis tool that has been used extensively to aid fuzzing~\cite{chen_2018_angoraefficientfuzzing,osterlund_2020_parmesansanitizerguidedgreybox,chen_2019_matryoshka,han_2019_synfuzz}.
DFSan instrumentation is enabled by compiling a binary with the option \texttt{-fsanitize=dataflow}, but additional effort is required to label input sources.
Inputs can be labeled by either an LLVM transformation pass, or manual modification of the source by adding the DFSan functions to create taint labels and assign them to sources of data appropriately.
Additionally, DFSan depends on a user-augmented ABI list in order to determine how to handle taint propagation when external library functions are called.
While dynamic data flow analysis is very powerful, it can also be quite brittle due to implicit data flows, the dependence on a complete ABI list, and complex I/O operations.

We use DFSan as a best-effort approach to aid the analyst in understanding the compartment list and how compartments might be unlocked.
The result of data flow analysis will label each compartment by the union of labels associated with the conditional executed immediately before the edge to the compartment entry block.
Currently, there are only two labels used by HM-fuzzing: (1) input label data and (2) harness label data.\footnote{In truth, there is a third category, which is \emph{unlabeled}, indicating the conditional is untainted by any currently known input, warranting investigation.}
As all benchmarks within FuzzBench, and also OSS-Fuzz, utilize the libFuzzer standard API calling convention, labeling the input data is done by the transformation pass that compiles the DFSan logging binary.
Labeling harness-controlled data is completed by the analyst, which normally requires minimal effort to add function calls on any options, flags, and default values that are initialized by the fuzzing harness.

\begin{table}
\centering
\footnotesize
\begin{tabularx}{\columnwidth}{Xlll}
\toprule
	\textbf{Benchmark} & \textbf{Missed Lines} & \textbf{Total Lines} & \textbf{\% Covered} \\
\midrule
  file          & 797   & 1,796 & 55.6\% \\
  freetype2     & 1,896 & 3,588 & 47.2\% \\
  libjpeg-turbo & 511   & 895   & 42.9\% \\
  libpcap       & 567   & 1,156 & 50.9\% \\
  libxml2       & 1,208 & 1,886 & 35.9\% \\
  proj4         & 567   & 896   & 36.7\% \\
\bottomrule
\end{tabularx}
  \caption{Source code coverage statistics for 10 functions with the most missed lines.  Assuming an analyst needed to review the top 10 functions with the most uncovered code, `Missed Lines' shows the total number of source code lines the analyst would be required to review.}%
  \label{table:missing_source_code_lines}
\end{table}

% Manually placed early
% \setlength{\tabcolsep}{4pt} % Slightly reduced padding for this table
\begin{table*}[tb]
    \footnotesize
    \caption[caption]{HM-fuzzing summary results analyzing 20 compartments identified after 6~hours of fuzzing.}%
  \label{table:summary_results}
\begin{tabularx}{\linewidth}{llXrrrcrrrr}
\toprule
                       &                  &               & \multirowcell{2}{\textbf{Stable}\\\textbf{Cmps.}} &  \multirowcell{2}{\textbf{Input}\\\textbf{Label}} & \multirowcell{2}{\textbf{Harness}\\\textbf{Labels}} & \multirowcell{2}{\textbf{Harness}\\\textbf{Mod?}} &                & \multirowcell{2}{\textbf{HM-Fuzzing}\\\textbf{Cov. $\Delta$}} & \multirowcell{2}{\textbf{Fn.}\\\textbf{Solns}} & \multirowcell{2}{\textbf{Edge}\\\textbf{Solns}}  \\
    \textbf{Benchmark} & \textbf{Version} & \textbf{SLOC} &                                                   &                                                   &                                                     &                                                   & \textbf{Solns} &                                                                   &                                                &                                                  \\
\midrule
 file          & v5.38 & 9K       &    17    &    0     &     0      &      N       &      3      &      +12\%       &       6 & 2        \\
 freeytype2    & v2.9 & 55K       &    17    &    2     &     1      &      N       &     11      &      +31\%       &       8 & 2        \\
 lcms          & v2.10 & 15K      &    16    &    0     &     6      &      Y      &      2      &      +14\%       &       1 & 1        \\
 libjpeg-turbo & v2.0.6 & 19K     &    14    &    11    &     0      &      Y      &      2      &       +0\%       &       0 & 0        \\
 libpcap       & v1.10.0 & 16K    &    14    &    13    &     0      &      N       &      7      &      +10\%       &       3 & 1        \\
 libxml2       & v2.9.8 & 95K     &    17    &    3     &     0      &      Y      &      6      &      +39\%       &       14 & 4       \\
 proj4         & v5.2.0 & 17K     &    17    &    2     &     0      &      N       &     16      &      +94\%       &       16 & 2       \\
\bottomrule
\multicolumn{11}{p{\linewidth}}{\footnotesize \textbf{Stable Cmps.}:   \# of compartments still locked after \textbf{96h of fuzzing}. A minimum of 14/20 compartments are stable despite much longer fuzzing campaigns.
 \textbf{Input Labels}:   \# of compartments where input data flow to the conditional blocking the compartment.
 \textbf{Input Labels}:   \# of compartments where input data flow to the conditional blocking the compartment.
 \textbf{Harness Labels}: \# of compartments where data labeled by harness flow to the conditional blocking the compartment.
 \textbf{Solns}:          \# of new seeds that unlock the entry block to one (or more) compartments.
 \textbf{HM-Fuzzing Coverage $\Delta$}: Code coverage improvement after 24h of HM-fuzzing vs.\ default.
 \textbf{Fn.\ Solns}: Number of functions with locked compartments that were all unlocked by HM-Fuzzing.
   % Number of function previously containing compartment entry block no longer appears in compartment list.}\\
\textbf{Edge Solns}: Number of functions with locked compartments where at least one compartment was unlocked, but not all.}\\
    %Function containing compartment entry block still appears; new entry block.}\\
\end{tabularx}
\end{table*}

\subsection{Human Intervention}%
\label{sub:human_intervention}

Before discussing the proposed role of the human in HM-Fuzzing, it is helpful to understand the human labor required without compartment analysis.
Based on community guidance,\footnote{Example from ``Visualizing Coverage'' in the libFuzzer tutorial https://github.com/google/fuzzing/blob/master/tutorial/libFuzzerTutorial.md} the recommended way to assess a fuzzing campaign is through a code coverage report.
% or "How good is my fuzzer?" https://llvm.org/docs/LibFuzzer.html
Unfortunately, although the HTML interface of a coverage report is easy to navigate, the overview of a moderately sized software project provides an overwhelming amount of information.
The overview contains the list of source files with various statistics per file such as function, line, or branch coverage.
Determining where to begin is not a simple task.
However, with a small amount of scripting it is possible to generate a list of functions that are sorted by the total number of uncovered source code lines.
Assuming the analyst starts with the function that contains the most uncovered lines, the next task is to determine where inside of that (potentially large) function the fuzzer is getting stuck.
% Figure~\ref{figure:coverage_report} shows a small excerpt from a coverage report of a partially-covered function in \texttt{file}.
The analyst must review various ``segments'' within the source code of the identified function: covered lines, uncovered lines, comments, and code not compiled due to compilation options.
Table~\ref{table:missing_source_code_lines} shows the number of lines an analyst would be required to review in order to begin to address missed coverage from the top 10 functions with the most uncovered source code lines.
For instance, in freetype2 an analyst would need to manually review up to 1,896 source code lines in order to determine how and where to intervene.

With compartment analysis, a security engineer or developer/maintainer can use the weighted list of compartments to prioritize human efforts to improve fuzzing coverage.
The list of compartments (see example in Table~\ref{table:example_file_compartment_list}) includes the results of the data flow analysis which hints how a compartment might be unlocked by input or harness modification.
At this point, the expertise and/or the level of familiarity the engineer has with the program under test will drive the next steps.
In some cases the symbol names provided in the compartment list might provide enough information for the engineer to find a solution seed.
The engineer might need to inspect the context of the code in the source and destination blocks of the edge leading to the locked compartment.
Using contextual information, the engineer may conclude that the provided seeds do not include a particular file format, or that a new fuzzing harness must be generated (see \S\ref{sub:case_study_1}), in order to cover exclusive conditional code.

\section{Implementation}%
\label{sec:implementation}

HM-fuzzing builds on LLVM's extensive transformation and analysis library.
The profiling data collection uses LLVM profile-guided optimization and source-based code coverage.
The data flow analysis uses LLVM DFSan to propagate labels for mutated input and harness controlled data.
The patch to \aflpp~is $\sim$200 lines of C to add an additional fork-server that executes every mutated input independent of, and in parallel to, any other enabled fork-servers.
The compartment analysis pass and the DFSan transformation pass are $\sim$800 and $\sim$400 lines of C++.
Additionally, we wrote Python scripts to execute the DFSan binary on a corpus of inputs and print the final list of compartments with annotations for data flow labels\cone~and any possible inputs that unlock a compartment.
These scripts allow an analyst to quickly evaluate candidate inputs~\ctwo~and identify those that represent solutions~\cthree.

% \begin{figure}
%     \includegraphics[width=0.9\columnwidth]{figures/fsmagic.c}
%     \caption{Example HTML coverage report.}%
%     \label{figure:coverage_report}
% \end{figure}

\section{Evaluation}%
\label{sec:evaluation}

\begin{figure*}[t]
    \centering
    \includegraphics[width=0.9\textwidth]{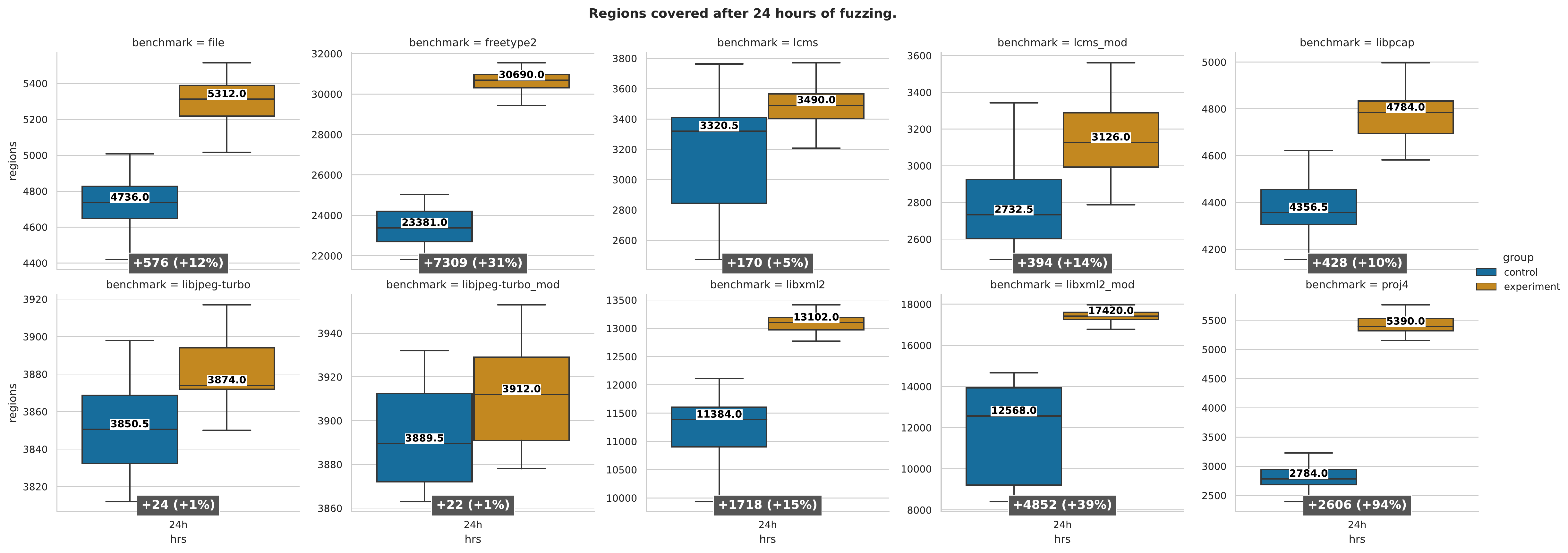}
    \caption{Regions covered for baseline fuzzing vs.~compartment analysis. Grey boxes show the absolute and relative increase in regions covered between the control and experiment.  The white boxes show the median regions covered after 50 trials.}%
    \label{figure:coverage-data}
\end{figure*}

To quantify the value of HM-fuzzing, we designed an experiment to measure the difference in coverage achieved when applying compartment analysis to a representative set of programs versus a baseline fuzzing campaign.
For this evaluation, we constructed a program test set consisting of seven programs, fuzzing harnesses, and seeds published as part of the Google FuzzBench corpus~\cite{google_fuzzbench}.
%Google FuzzBench, which is based on OSS-Fuzz~\cite{google_2020_ossfuzz}, provides an open source corpus of programs, test harnesses, and seeds, and therefore represents an ideal platform for reproducible experimentation.
The particular programs we selected as part of the test set represent a variety of input file formats of varying complexity, and include both text and binary input formats.  Table~\ref{table:summary_results} provides details on each of these selected benchmarks.

We selected \aflpp~\cite{fioraldi_2020_aflcombiningincremental} as a state-of-the-art example of a grey-box mutational fuzzer.  AFL++ is a fork of AFL that incorporates a number of community and academic improvements (both AFL and non-AFL-based), and consistently appears at or near the top of the Google FuzzBench rankings.

Our experiment was structured to simulate the workflow of a real-world fuzzing campaign.
For the control group, we ran AFL++ for 24~hours without intervention using unmodified fuzzing harnesses and seed sets obtained from FuzzBench.
In the experimental group, we ran AFL++ for 12~hours, paused fuzzing, and conducted a compartment analysis on the incremental coverage obtained at that point using the 20~heaviest hitters reported by the analysis.
We then continued fuzzing for 12~more hours.
Each experiment group was repeated 50~times in accordance with community guidance on testing randomized algorithms~\cite{klees_2018_evaluatingfuzztesting}.
All experiments were conducted on servers with Intel Xeon E5-2670 2.5~GHz CPUs and 256~GB RAM with Ubuntu~16.04 as the host operating system.

Compartment analysis was conducted by the authors on the benchmarks after fuzzing for 12~hours to collect the required profiling data.
Approximately 8~hours were spent reviewing the top~20 compartments on each benchmark.
The compartments were reviewed in a greedy, best-effort approach from greatest weight to least.
When it was determined that a compartment required an additional input, these inputs were found by searching for a particular file format or by manually crafting new inputs based on other inputs.
Only minor modifications were made to the fuzzing harnesses, such as altering options or flag values, so that coverage comparisons could be made with the default harness. % Trying to say we did not alter the harness to call the library API via a different method.

%In the remainder of this section, we describe the overall results of this experiment~(\S\ref{sub:compartment_analysis_overview}) before discussing four case studies in greater detail~(\S\ref{sub:case_study_0}-\ref{sub:case_study_2}).

\subsection{Compartment Analysis vs.~Baseline}%
\label{sub:compartment_analysis_overview}

A summary overview of our experimental results are shown in Table~\ref{table:summary_results}, while Figure~\ref{figure:coverage-data} presents boxplots comparing coverage attained by the baseline (blue) and experimental (yellow) runs.  Median values are annotated on the boxplots as well as the raw and percentage increase in median coverage at the bottom of each plot.

Overall, all targets except for libjpeg-turbo showed significant coverage gains when compartment analysis was applied.
These gains ranged from moderate---10\% in the case of libpcap---to dramatic---94\% in the case of proj4.
proj4 exhibited the greatest number of solutions, where 16~interventions resulted in complete coverage of a reported blocking edge's parent function.
There is an observable positive correlation in the table between the number of solutions and the amount of new coverage achieved.
However, two benchmarks attained moderate increases in coverage (10-14\%) even with a low unlock rate (2-3).
This demonstrates that even a small number of successful interventions can still yield significant coverage gains.
Despite the inherent randomness of fuzzing, top 20 compartments are mostly stable from 12 hours through 96 hours of fuzzing.
As identified by the analyses described in Section~\ref{sub:optimal_edge_ranking}, between 14 and 17 of the top 20 compartments are the same for the nine trials performed.

\subsection{Case Study: Harness Modifications}%
\label{sub:case_study_0}

There were varied sources for the coverage improvements represented in the results.  For instance, harness taint propagated to conditionals blocking compartments for both lcms and freetype2.  In the former case, it was quickly discerned that the fuzzing harness's use of default values for the \ccode{Intent} and \ccode{flags} variables were directly responsible for blocking access to two compartments.  To address this, we modified the harness for lcms to use different options.  The
difference in coverage achieved with the lcms harness with and without modifications to remove access those two compartments is shown in the boxplots labeled \textsf{lcms} and \textsf{lcms\_mod}, respectively, in Figure~\ref{figure:coverage-data}.  Interestingly, median coverage actually \emph{decreased} with the modified harness from 3320 to 2732 regions (-18\%).  In investigating this, it became clear that both sets of options should be explored by the fuzzer in order to obtain the best
coverage.  The union of the coverage between both the unmodified and modified lcms harnesses was 4104 regions, an increase of 24\% over the unmodified harness.  In fact, this observation suggests that, as a rule, it may be fruitful to add library options (or, more broadly, program arguments and environment variables) to the scope of mutated inputs to programs under test to systematically reduce this source of uncoverable code.\footnote{Of course, mutating API parameters must be done in such a
way as to preserve valid uses of the API\@.  For instance, crashes resulting from fuzzing a pointer argument are unlikely to be useful findings.}

In freetype2's case, we found one compartment with harness-labeled data influencing the blocking conditional.  However, given the analyst's time budget it was determined that input set modifications would be a better use of effort.  This determination was due to the presence of multiple compartment symbols referencing font file formats that fuzzers were unable to mutationally synthesize.  The remainder of the interventions performed during compartment analysis took the form of input set augmentation.

\begin{figure}[tb]
  \centering
  \includegraphics[width=0.9\linewidth]{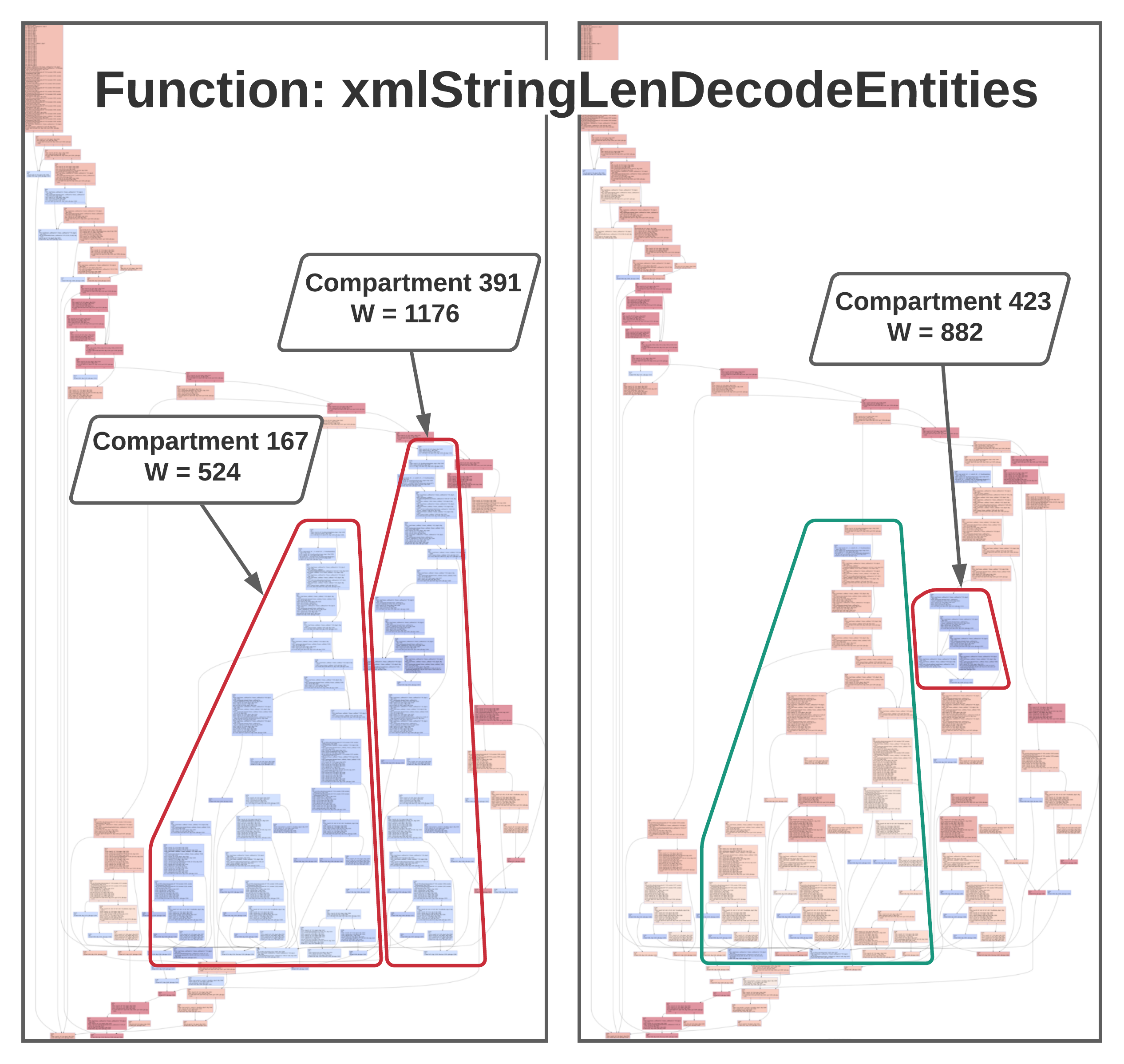}
  \caption{Compartments within the function \texttt{xmlStringLenDecodeEntities}. Initially, two uncovered compartments ($167$ and $391$) are identified as shown on the left. After human intervention, fuzzing unlocks both compartments, but a smaller sub-compartment ($423$) remains locked.}%
  \label{figure:xmlStringLenDecodeEntities}
\end{figure}

\begin{figure*}
  \centering
  \includegraphics[width=0.9\textwidth]{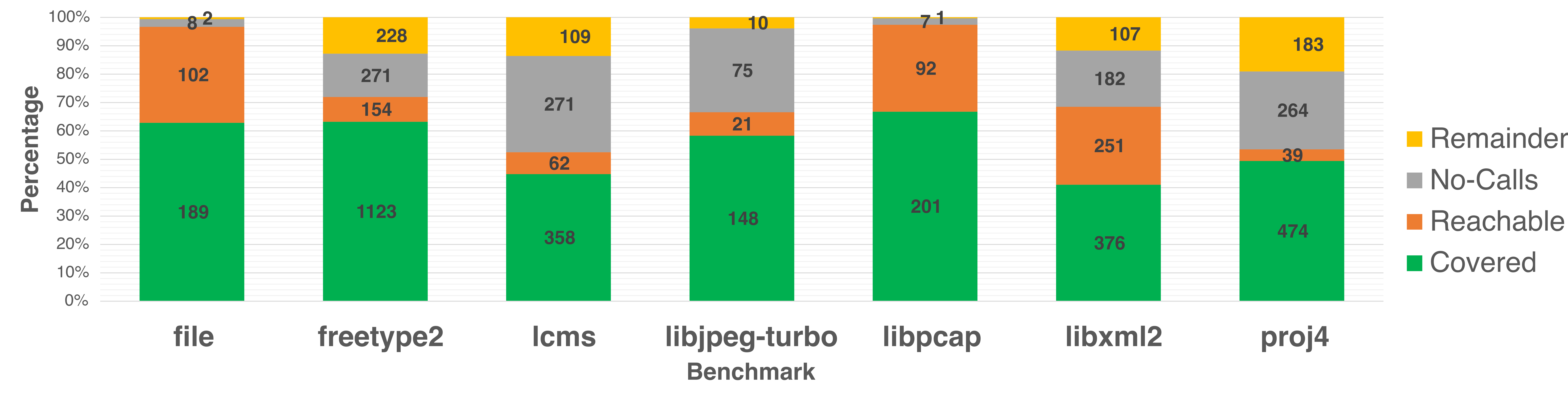}
  \caption{Function coverage after 6~hours and the static analysis of uncovered functions. (1)~Reachable means there is a known path through the ICFG to reach the uncovered function. (2)~No-Calls means the function has no known call-site.  (3)~Remainder are functions that have call-sites, but the path from the entrypoint is undetermined.}%
  \label{figure:FunctionCoverage}
\end{figure*}

\subsection{Case Study: libxml2}%
\label{sub:case_study_1}

In libxml2, our prototype identified 20~uncovered compartments ranked by potential coverage impact.
By examining our analysis results, we determined that 5~new compartments could be unlocked by adding 4~new, simple seeds.
Additionally, we determined that the harness should be modified to enable the \texttt{SAX1} option (as shown previously in Listing~\ref{lst:example_libxml2}).
This modification, combined with two more SAX1-based seeds, unlocked an additional 12~compartments.

\begin{listing}[tb]
  \begin{minted}[fontsize=\footnotesize]{c}
else if ((c == '&') && (what & XML_SUBSTITUTE_REF))
  \end{minted}
    \caption{Source code guarding libxml2 compartment 167 (compartment shown in Figure~\ref{figure:xmlStringLenDecodeEntities}).}%
  \label{lst:libxml2_167_code}
\end{listing}

These changes resulted in a 39\% increase in the median code regions covered when combining the new seeds and the modified fuzzing harness.
Two of the compartments are shown in Figure~\ref{figure:xmlStringLenDecodeEntities}, which both exist in the function \texttt{xmlStringLenDecodeEntities}.
The left compartment in the figure denoted by the entry basic block \texttt{167} was unlocked by an additional seed, but the compartment on the right (\texttt{391}) was only partly unlocked.
The weight of compartment \texttt{167} is composed of 257~unexecuted instructions dominated by the entry block as well as 267~instructions belonging to the function \texttt{xmlParseStringEntityRef} which is uniquely called by \texttt{xmlStringLenDecodeEntities}.
This edge is represented in the source code shown in Listing~\ref{lst:libxml2_167_code} which checks for the character \texttt{'\&'} inside an entity which prefixes a formerly defined entity.
This input demonstrates slightly more complex constraints as the entity must first have a valid definition, and subsequently have a valid reference.
In this case, AFL++ was able to generate XML inputs that contained valid entities, but not a sequence where a valid entity \(e_1\) is followed by another entity \(e_2\) that refers to \(e_1\).  An example input that satisfies these constraints is shown in Listing~\ref{lst:libxml2_167_input}.

\begin{listing}[tb]
\begin{minted}[fontsize=\footnotesize]{xml}
<!ENTITY beta "&#946;">
<!ENTITY ab "&#947;&beta;">
  \end{minted}
    \caption{Input to to unlock libxml2 compartment 167. The variable \ccode{beta} is defined, then referenced.}%
  \label{lst:libxml2_167_input}
\end{listing}

%% 10page
% \begin{figure}[t]
%   \includegraphics[width=1.0\linewidth]{figures/fuzz-both-input}
%   \caption{Input format for libpcap harness \texttt{fuzz\_both}. The size field determines the length of the filter string (unstructured data) and the start of the pcap file (structured data), which is a brittle structure to mutate randomly.}%
%   \label{figure:fuzz-both-input}
% \end{figure}

An additional compartment (\texttt{391}) could not be fully unlocked by modifying the harness or providing additional seeds.
Fully unlocking this compartment requires referencing an external entity---such as another file on the file system---and then successfully loading that external entity.
With new seeds, the fuzzer was able to make some progress by reducing the locked compartment from \texttt{391} with a weight of 1176 unexecuted instructions to compartment \texttt{423} with weight 882.  This uncovered code belongs to the unexecuted function \texttt{xmlLoadEntityContent}, which is uniquely called by \texttt{xmlStringLenDecodeEntities}.

\subsection{Case Study: proj4}%
\label{sub:case_study_2}

The compartment analysis of proj4 demonstrates the potential effectiveness of indirect call analysis.
On the first iteration of compartment analysis, indirect call targets (i.e., functions with no static call references in the source code) were not included.
When indirect call targets were added in the compartment analysis, 14~of the top~20 compartments from the initial analysis were replaced with new compartments whose entry block is the function entry of an indirect call target.
The fraction of uncovered code reachable only through indirect calls is reflected in Figure~\ref{figure:FunctionCoverage}.  As this figure shows, 39~functions are reachable through static code analysis while 264~functions have no corresponding call sites.
The majority of the uncovered functions (``No-Calls'') are merely the 154~possible cartographic projections that proj4 is capable of converting.
Even though the mutations required to reach these projections are rather simple, AFL++ was unable to cover these functions after \textbf{96~hours of fuzzing}.
Using the compartment analysis including indirect call targets, solution inputs were easily generated using the function names as an indicator of the required input projection description.  For instance, \texttt{+proj=omerc} indicates that conversions to/from the Oblique Mercator projection are required.

%% 10page
% \subsection{Case Study: libpcap}%
% \label{sub:case_study_3}
% 
% libpcap's compartment analysis identifies 15~compartments that all reside within the same function (\ccode{pcap_parse}).  In fact, each of these compartments has the same source location as they are all part of one switch statement.
% This switch statement implements various options for the \texttt{PCAP-FILTER} syntax and functionality.\footnote{\url{https://www.tcpdump.org/manpages/pcap-filter.7.html}}
% The fuzzing harness used in the FuzzBench libpcap benchmark attempts to fuzz both the packet filter syntax plus related functionality \emph{and} the packet processing code using one harness driven by a single input buffer.
% What compartment analysis clearly reveals in this case is that AFL++ cannot effectively fuzz both the filter syntax options and the packet processing with the default harness and chosen seeds.
% 
% The inputs that unlock these compartments are shown in Figure~\ref{figure:fuzz-both-input}.  The first byte of the inputs determine the length of the filter string.  This length is immediately followed by the filter string itself, to which the pcap file is appended.
% These inputs were crafted manually by creating separate valid filter options and pcap files.
% 
% We note that the libpcap project within OSS-Fuzz contains alternate fuzzing harnesses that fuzz the filter options and the packet processing separately.  Compartment analysis highlights that this approach to fuzzing libpcap is strictly better than FuzzBench's combined approach when it comes to maximizing coverage.

\subsection{Fuzz Duration vs HM-fuzzing}%
\label{sub:fuzz-duration}

\begin{figure*}
    \centering
    \includegraphics[width=0.95\textwidth]{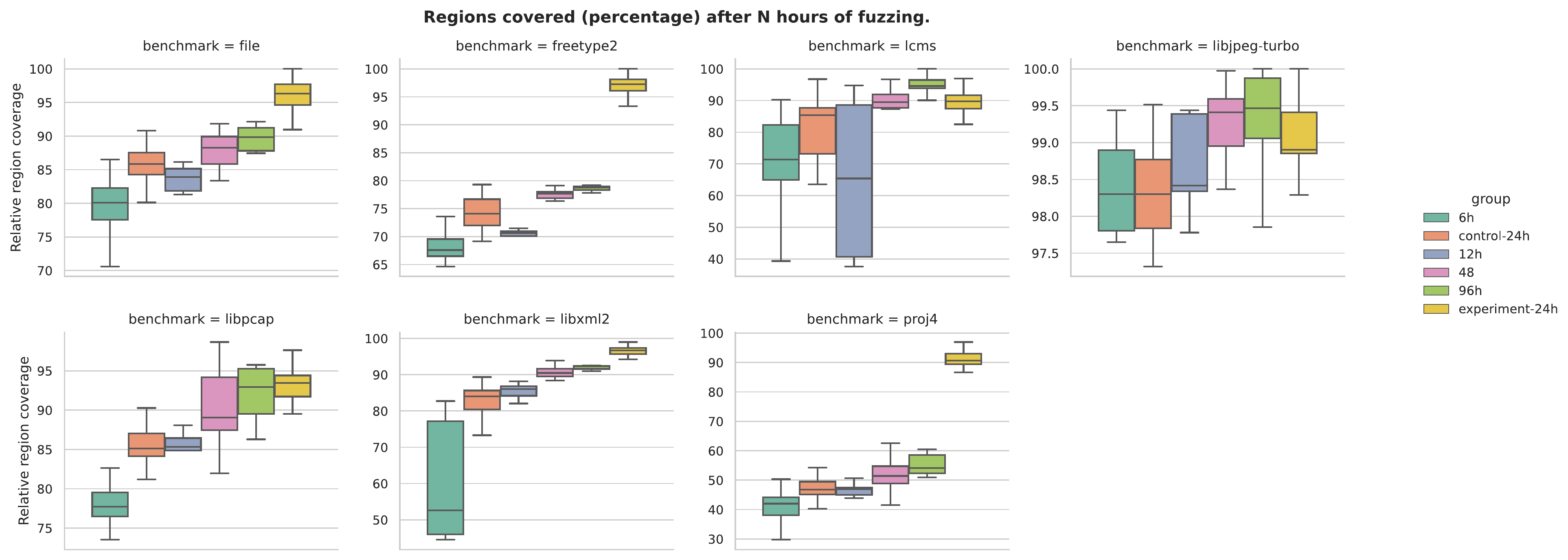}
    \caption{Fuzzing longer versus intervening with HM-fuzzing.  In 5 out of 7 benchmarks performing compartment analysis was more productive than fuzzing for 96 hours.}%
    \label{figure:fuzz-duration}
\end{figure*}

We conducted additional fuzzing campaigns with constant parameters (seeds, binaries) but varied durations: 6h, 12h, 48h, and 96h. The relative program coverage across the various durations is shown for each benchmark in Figure~\ref{figure:fuzz-duration}.
These alternate experiments demonstrate that HM-fuzzing for 24 hours often performs at least as well as a baseline fuzzer would after 96 hours.
The results closely parallel the successes reported in Figure~\ref{figure:coverage-data}.  For the file, freetype2, libpcap, libxml2, and proj4 benchmarks it is more beneficial to use compartment analysis during a 24 hour campaign than it would be to fuzz for 96 hours.
For the other benchmarks (lcms and libjpeg-turbo), conducting compartment analysis during a 24 hour campaign provides comparable results to fuzzing for 48 hours without intervention. Furthermore, it is only marginally better ($\sim$1\% increase) to fuzz for the full 96 hours without compartment analysis than it is to fuzz for 24 hours with the compartment analysis.

\section{Discussion}%
\label{sec:discussion}

While we believe that HM-fuzzing can make a valuable contribution to real-world fuzzing workflows by significantly broadening coverage of programs under test, the technique is not a panacea and has some limitations that we discuss in the following.

\paragraph{Efficacy scales with human expertise}%
\label{par:efficacy_scales_with_human_expertise_}

The entire point of compartment analysis is to efficiently integrate human expertise into the fuzzing loop in such a way that maximizes that human's time---i.e., the program points where interventions might have the biggest impact are the ones that are presented to the security analyst.  However, identifying the proper intervention to make given the information at hand is left to the analyst.  While our experience suggests that there is often ample information available---whether in the form of symbol names, or compartment analysis hints such as taint-based attribution of harness limitations---interpreting this information may be gated by the analyst's familiarity with the programs under test in terms of their purpose, design, implementation, and expected inputs.  We believe it is reasonable to expect some modicum of familiarity in cases where developers engage in fuzzing as a component of their testing process such as SAGE~\cite{godefroid_2012_sagewhiteboxfuzzing} or ClusterFuzz~\cite{google_2020_clusterfuzz}; or, alternatively, where security analysts coordinate long-term fuzzing campaigns against a fixed set of targets as in the case of OSS-Fuzz~\cite{google_2020_ossfuzz} or FuzzBench~\cite{google_fuzzbench}.
%Ultimately, however, only wider experimentation will be able to validate this hypothesis.

\paragraph{Some programs are simply hard targets}%
\label{par:some_programs_are_simply_hard_targets_}

% Unfortunately, a degree of familiarity with a test program is not always sufficient to formulate informed interventions.
In our experience with this technique to date, the further a program strays from the mean path of general computer science knowledge, the more difficult it will be for the average analyst to quickly identify appropriate input set additions or harness modifications.

Consider the case of libjpeg-turbo, the test program where compartment analysis was least effective.  The heaviest hitters in the compartment analysis table were located in functions responsible for image decoding, e.g., \ccode{jinit_master_decompress}.  In particular, it seemed that harness modifications to the library parameters controlling how the discrete cosine transform would be performed when decoding images would unlock these compartments.  However, despite investing approximately 16 hours in pursuit of these compartments, significant advances were not made on this target.  While the analyst that carried out this experiment is not an expert on JPEG image decoding, it is also not reasonable to expect that the average analyst is either.  Thus, there are clearly some targets that are not well-suited to compartment analysis.
In fact, this experience seems to align well with coverage data from FuzzBench, which shows that automated fuzzers also have difficulty with the complex path constraints found in libjpeg-turbo.

Contrast libjpeg-turbo with the file utility.
Among the heaviest hitters in the compartment analysis table are symbols such as \ccode{file_trycdf}, \ccode{file_zmagic}, \ccode{file_tryelf}, and \ccode{file_fsmagic}.
These names would immediately suggest to many security analysts the exact kinds of input files that are involved: respectively, a Microsoft Word document, a compressed file, an ELF file, or a file system image.
%Further inspection of the code corresponding to the compartment entry point refines this further, indicating either particular features that need to be present in these files or whether harness modifications are required.
In this respect, file certainly represents a better candidate for compartment analysis than libjpeg-turbo for the average analyst.

%% 10page
% \paragraph{Why didn't you find any new bugs?}%
% \label{par:why_didn_t_you_find_any_new_bugs_}
% 
% No previously unknown bugs were uncovered in the course of our experiments.  A natural question to ask is why that is the case, if compartment analysis is supposed to significantly broaden coverage.  Should this not lead to a non-zero bug discovery rate?
% 
% We believe the answer lies in the program test set selection methodology.
% The targets we selected were drawn from FuzzBench, which in turn is primarily an early snapshot of selected projects within of OSS-Fuzz.
% The motivation for using this corpus is because of its openness for replication's sake and recognition as a benchmark standard fuzz testing framework.
% However, Google has been continually fuzzing the OSS-Fuzz corpus, and thus FuzzBench targets, for years.
% The project (benchmark) maintainers have been conducting their own versions of HM-fuzzing by creating new harnesses, improving existing ones and adding to the seed corpus as new bugs are found.
% Therefore, it is difficult to discover new bugs in this corpus, certainly within the bounded scope of experiments to support this paper.

\section{Related Work}%
\label{sec:related_work}

HM-fuzzing builds on an extensive literature in improving fuzz testing.  Since the introduction~\cite{miller_1990_empiricalstudyreliability} of fuzzing, improvements have steadily been attained in a number of ways.  For instance, it has long been recognized that systematic approaches for seed selection~\cite{rebert_2014_optimizingseedselection} and maintenance of the input queue~\cite{woo_2013_schedulingblackboxmutational} is important to broaden coverage and minimize wasted trials.  More sophisticated fuzz configuration scheduling algorithms have been proposed to prioritize inputs that are more likely to produce new coverage~\cite{bohme_2016_coveragebasedgreyboxfuzzing} or to drive execution towards particular points in a program under test~\cite{bohme_2017_directedgreyboxfuzzing,osterlund_2020_parmesansanitizerguidedgreybox}.  Prior work has investigated new approaches for mutation operator scheduling in order to more efficiently cover test programs~\cite{lyu_2019_moptoptimizedmutation}.  A particularly active area of work over the past decade has investigated linking input bytes to the coverage frontier in order to focus mutations and attain coverage more efficiently~\cite{leek_2007_coveragemaximizationusing,wang_2010_taintscopechecksumawaredirected,rawat_2017_vuzzerapplicationawareevolutionary,chen_2018_angoraefficientfuzzing,aschermann_2019_redqueenfuzzinginputtostate}.  Hybrid fuzzing that leverages concolic execution to assist a conventional mutational fuzzer has been extensively investigated~\cite{majumdar2007hybrid,pak2012hybrid,stephens_2016_drilleraugmentingfuzzing,yun_2018_qsympracticalconcolic,zhao_2019_sendhardestproblems}.

All of the above work along these different axes is complementary to compartment analysis.  Since state-of-the-art fuzzers are currently unable to achieve anything close to full coverage for non-trivial targets~\cite{klees_2018_evaluatingfuzztesting,bundt_2021_evaluatingsyntheticbugs,bohme_2020_fuzzingexponentialcost}, compartment analysis can be used to assist any fuzzer in reaching code that it would otherwise not cover within a given resource budget once it has stopped achieving new coverage on its own.

\paragraph{Human-in-the-loop analyses}%
\label{par:human_in_the_loop}

More directly related to compartment analysis are other efforts at tightly integrating humans into an otherwise automated analysis.  One prominent example is HaCRS~\cite{shoshitaishvili_2017_risehacrsaugmenting}, which crowdsources inputs to a program in an attempt to achieve a variety of tasks that includes broadening coverage.
%These inputs can be suggested on the user's own volition, or with the help of ``symbolic tokens'' suggested by a concolic execution system.
While the high-level goals of HaCRS and compartment analysis are similar, HaCRS does not algorithmically define how guidance is suggested.
%Rather, this is opaque and implementation-dependent on the underlying analysis framework, which in practice is angr~\cite{shoshitaishvili_2016_sokstateart}.  In contrast, compartment analysis is strictly defined and focused on solving a well-defined problem, namely that of providing new inputs or harness modifications that correspond to conditional edges that gate large amounts of uncovered code.

Another closely related effort is Maier et al.'s JMPscare~\cite{maier_2021_jmpscareintrospectionbinaryonly}, which analyzes the cumulative coverage of a fuzzing campaign to identify the coverage frontier.
%The system supports patching target binaries in order to then use forced execution to cover new code across this frontier.  While there are some similarities in the techniques used,
JMPscare relies on forced execution to cover new code which is inherently unsound.  In contrast, conclusions drawn from compartment analysis are always true.
%More broadly, JMPscare is oriented towards exploratory analyses, where fuzzing is used to assist with a primarily manual analysis.  Compartment analysis adopts the converse model, where manual analysis is used to provide occasional assistance to a mostly automated fuzzing campaign.

Aside from fuzzing, manual assistance has also been used to help improve symbolic execution. Bornholt et al.~\cite{bornholt_2018_explodes} design a system to identify the cause of performance bottlenecks in symbolic evaluation tools; the authors then manually intervene to help resolve the bottlenecks. More directly related to bug-finding, Galea et al.~\cite{galea_2019_hemiptera} created a custom bug benchmark, Hemiptera, with 130 bugs in utilities such as file, libjpeg-turbo, and tcpdump, and evaluated the impact of manual intervention when attempting to find these bugs with KLEE~\cite{cadar_2008_kleeunassistedautomatic}. In this case, both the identification of bottlenecks and the interventions were performed manually.

\paragraph{Automated harness generation}%
\label{par:automated_harness_generation}

Related to the idea of harness modifications as used in compartment analysis is automated harness generation.
%In particular, this area focuses on libFuzzer-style testing, where some manual effort must be expended to write a fuzzing harness and testing is only valid if the API under test is used correctly (or at least as real programs use it).
FUDGE~\cite{babic_2019_fudgefuzzdriver} analyzes source code to extract valid sequences of API calls.  FuzzGen~\cite{ispoglou_2020_fuzzgenautomaticfuzzer} observes whole system behavior to collect a diverse set of valid ways in which an (open source) API can be used.  Winnie~\cite{jung_2021_winniefuzzingwindows} harnesses closed-source Windows applications by dynamically analyzing targets to find and fuzz functions.
%As with other fuzzing techniques,
These efforts are complementary to our own, and can be useful in bootstrapping harnesses for new corpora.

\section{Conclusions}%
\label{sec:conclusions}

In this paper, we introduced HM-fuzzing and compartment analysis which integrates targeted human guidance into fuzzing workflows. %, and has the express aim of surmounting the ``coverage barrier'' that stymies automated fuzzers.
Compartment analysis uses a combination of static interprocedural control-flow analysis, dynamic data-flow analysis, and a novel weighting function on incremental coverage to present security analysts with a prioritized list of human intervention points.  These interventions, which can take the form of additions to the input queue or harness modifications, allow human-guided fuzzers to achieve significantly greater coverage than automated fuzzing alone in almost all cases considered in our evaluation.

In future work, we plan to explore other fruitful mechanisms for integrating judicious human guidance into bug finding.  While automated techniques will inevitably become ever more efficient at finding software vulnerabilities, we believe there will always be code that is too difficult to reach with automation alone.  The human in the machine, \emph{homo in machina}, can help bridge this gap.

\thanks{\scriptsize{%
\noindent
DISTRIBUTION STATEMENT A.
Approved for public release. Distribution is unlimited. This material is based upon work supported by the Under Secretary of Defense for Research and Engineering under Air Force Contract No. FA8702-15-D-0001. Any opinions, findings, conclusions or recommendations expressed in this material are those of the author(s) and do not necessarily reflect the views of the Under Secretary of Defense for Research and Engineering. \copyright\the\year{} Massachusetts Institute of Technology. Delivered to the U.S. Government with Unlimited Rights, as defined in DFARS Part 252.227-7013 or 7014 (Feb 2014). Notwithstanding any copyright notice, U.S. Government rights in this work are defined by DFARS 252.227-7013 or DFARS 252.227-7014 as detailed above. Use of this work other than as specifically authorized by the U.S. Government may violate any copyrights that exist in this work.%
\endgraf
}}

\printbibliography

\end{document}